# High-entropy perovskites as new photocatalysts for cocatalyst-free water splitting

Ho Truong Nam Hai[a,b], Makoto Arita[c], Kaveh Edalati[a,b,d],*

[a] WPI, International Institute for Carbon Neutral Energy Research (WPI-I2CNER), Kyushu University, Fukuoka 819-0395, Japan

[b] Department of Automotive Science, Graduate School of Integrated Frontier Sciences, Kyushu University, Fukuoka 819-0395, Japan

[c] Department of Materials Science and Engineering, Faculty of Engineering, Kyushu University, Fukuoka 819-0395, Japan

[d] Mitsui Chemicals, Inc. - Carbon Neutral Research Center (MCI-CNRC), Kyushu University, Fukuoka, 819-0395, Japan

*Corresponding author (E-mail: kaveh.edalati@kyudai.jp; Phone: +81 92 802 6744)

**Abstract**

The photocatalytic water-splitting process is thermodynamically challenging and requires catalysts with suitable band structures, as well as the presence of supporting cocatalysts. By considering the unique charge carrier mobility in perovskites, this study introduces three new $ABO_3$-type high-entropy perovskites $(Ba_{1/2}Sr_{1/2})(Ti_{1/3}Zr_{1/3}Hf_{1/3})O_3$, $(Ba_{1/2}Sr_{1/2})(Ga_{1/3}In_{1/3}Sn_{1/3})O_{3-\delta}$ and $(Ba_{1/2}Sr_{1/2})(Ti_{1/3}Zr_{1/3}Sn_{1/3})O_3$ for cocatalyst-free photocatalysis. The three catalysts, having a single-phase cubic structure, are designed by considering configurational entropy, tolerance factor, octahedral factor, ionic radius deviation and valence deviation of >1.5$R$ ($R$: gas constant), 0.9-1.0, 0.4-0.8, >0.3 and >0.3, respectively. The perovskites exhibit similar valence band tops, while their bandgaps vary slightly depending on the composition at the B-site (slightly lower bandgap by including $d^{10}$ cations). Additionally, all three materials demonstrate effective hydrogen generation without the need for added cocatalysts. This investigation confirms that high-entropy oxide perovskites can offer significant potential for cocatalyst-free photocatalytic reactions.

**Keywords:** High-entropy oxides (HEOs); Hydrogen ($H_2$) production; Photocatalysis; High-entropy photocatalysts; Co-catalyst

1. **Introduction**

The depletion of fossil fuels such as petroleum and coal has recently driven the development of new technologies capable of producing alternative fuels with high energy efficiency. Hydrogen ($H_2$) is recognized as a clean energy carrier with high energy density, and it releases energy and water as the only byproduct by combustion or electrochemical reaction, thus forming a sustainable and infinite energy cycle [1]. While commercial steam reforming is not a clean $H_2$ production method, photocatalytic water splitting stands out as a clean technology that utilizes solar energy and catalysts to reduce water and produce green $H_2$ [2]. In a water-splitting process, the excitation of electrons with input energies higher than the bandgap of the material plays a crucial role in initiating redox reactions [3]. However, this process faces significant thermodynamic challenges, and is greatly affected by the separation, migration and recombination of electrons and holes [4]. Moreover, to effectively harness solar energy, catalysts must possess an appropriate bandgap (around 1.2-3.0 eV) [5] for photocatalytic applications such as water splitting [2,6–10], $CO_2$ conversion [11], nitrogen fixation [12] and waste oxidation [13,14]. Currently, the efficiency of solar-to-hydrogen (STH) conversion achieved through photocatalytic water splitting is only about 1-2 % [15], while increasing the efficiency to 10 % could meet one-third of global energy demand [16].

To enhance the water-splitting efficiency for $H_2$ evolution, the use of cocatalysts (such as platinum, palladium, ruthenium, etc.) is essential, as they enhance charge separation efficiency and reduce electron-hole recombination by providing electron-donor sites [4]. However, cocatalysts also present several limitations. First, differences in lattice structure between the catalyst and the cocatalyst can result in significant energy loss and hinder charge separation at their interface [17]. Second, some cocatalysts are being photo-corrosive under high-intensity and long-term light illumination [18]. Corroded catalysts can sometimes trigger reverse catalytic reactions, thereby reducing $H_2$ generation efficiency [19]. Finally, the use of cocatalysts limits scalability and practical application due to their high cost and inconsistent effectiveness across different types of cocatalysts [17]. Therefore, an effective solution is to design a catalyst with an appropriate bandgap that ensures high performance and stability without the need for a cocatalyst for $H_2$ evolution.

Perovskites are known as a common group of compounds with the general chemical formula $ABX_3$ [20], which may offer a solution for cocatalyst-free photocatalysis. In this structure, the A site contains alkali metals, alkaline earth metals, or rare earth metals; the B site is occupied by transition metals or post-transition metals; and the X site contains oxygen or halogens [21]. In the ideal perovskite crystal lattice, each B atom has six-fold coordination and bonds with six X atoms to form an octahedral structure [22]. Meanwhile, each A atom with twelve-fold coordination is arranged at the center of the voids of the cuboctahedral cavity [22]. The stability of the structure and the diversity in electron movement and properties through substitution at the A and B sites make perovskites suitable for catalytic reactions [23]. Such substitutions can be particularly achieved by using the high-entropy material concept. With the rapid advancement of high-entropy alloys and ceramics, the notion of high-entropy perovskites (HEPs) was also introduced, although their significance has not been widely recognized. High-entropy ceramics, including high-entropy oxides and perovskites, are composed of at least five principal cations with a configuration entropy larger than or similar to $1.5R$ ($R$: gas constant) [24,25]. With enhanced thermodynamic stability and the emergence of new properties due to the cocktail effect, high-entropy materials show more diverse potential in photocatalytic applications compared to traditional catalysts [24,25]. Despite growing interest in high-entropy photocatalysts, few studies have examined HEPs for cocatalyst-free $H_2$ production.

In this study, three single-phase HEPs with the general configuration $(Ba_{1/2}Sr_{1/2})[(B_a)_{1/3}(B_b)_{1/3}(B_c)_{1/3}]O_3$, where combination of $B_a$, $B_b$ and $B_c$ are varied from the only $d^0$ group (Ti, Zr, Hf) to $d^{10}$ group (Ga, In, Sn) and $d^0+d^{10}$ group (Ti, Zr, Sn), are successfully synthesized for cocatalyst-free photocatalysis. Although perovskites are mainly synthesized by sol-gel [26], ball-milling [27], hydrothermal [28], co-precipitation [29] spark-plasma sintering [30] and pulsed laser deposition [31] methods, high-pressure torsion (HPT) was used for the synthesis in this study to have an ideal compositional homogeneity in the form of single phases [32]. Photocatalytic water splitting experiments demonstrate that these new materials can generate $H_2$ without the support of a platinum cocatalyst, revealing the potential of HEPs for practical applications.

## 2. Materials and methods
### 2.1. Catalyst design

To determine the theoretical basis for the design of new HEPs, parameters such as configuration entropy ($\Delta S_{config}$), tolerance factor ($t$), octahedral factor ($\mu$), octahedral mismatch ($\Delta\mu$), ion radius deviation ($\delta(r)$) and valence deviation ($\delta(V)$) of the synthesized materials need to be considered. These values can be calculated based on Eqs. 1-5 for a perovskite structure with an $ABX_3$ configuration [22,24,25].

$$\Delta S_{\text{config}} = \Delta S_{\text{A,config}} + \Delta S_{\text{B,config}} = -R \sum_{i=1}^{N} x_i \ln x_i \qquad (1)$$

$$t = \frac{\bar{r}_A + r_X}{\sqrt{2}(\bar{r}_B + r_X)} \qquad (2)$$

$$\mu = \frac{\sum_{j=1}^{M} r_{Bj}}{3r_X} \quad (3)$$

$$\delta(r) = \sqrt{\sum_{i=1}^{N} x_i \left[\frac{r_i - \bar{r}}{\bar{r}}\right]^2} \quad (4)$$

$$\delta(V) = \sqrt{\sum_{i=1}^{N} x_i \left[\frac{V_i - \bar{V}}{\bar{V}}\right]^2} \quad (5)$$

In these equations, $R$ is the gas constant, $x_i$ is the atomic fraction of the $i$th cation, $N$ is the number of cations, $\bar{r}_A$, $\bar{r}_B$ and $r_X$ are the average ionic radius of A-type atoms, B-type atoms and anion, $r_{Bj}$ is the ionic radius of the $j$th B-type element, $M$ is the number of B-type elements, $r_i$ is the ionic radius of the $i$th cation, $\bar{r}$ is the average ionic radius of all cations, $V_i$ is the valence of the $i$th cation and $\bar{V}$ is the average valence of all cations. To have a HEP, (i) $\Delta S_{config}$ should be higher than 1.5$R$, (ii) $t$ should be between 0.9 and 1.0, (iii) $\mu$ should be between 0.4 to 0.8, (iv) $\delta(r)$ should be over 0.3, and (v) $\delta(V)$ should be larger than 0.3 [22,24,33–40]. In this study, two elements (Ba, Sr) were selected for the A site and six elements with either $d^0$ (Ti, Zr, Hf) or $d^{10}$ (Ga, In, Sn) electronic configuration were selected for the B site to produce HEPs with a formula $(Ba_{1/2}Sr_{1/2})[(B_a)_{1/3}(B_b)_{1/3}(B_c)_{1/3}]O_3$. All 20 compositions in this formula satisfy the requirement for HEPs. For example, Fig. 1 examines the tolerance factor and octahedral factor of all possible HEPs, showing that all combinations can produce HEPs with an ideal cubic structure.

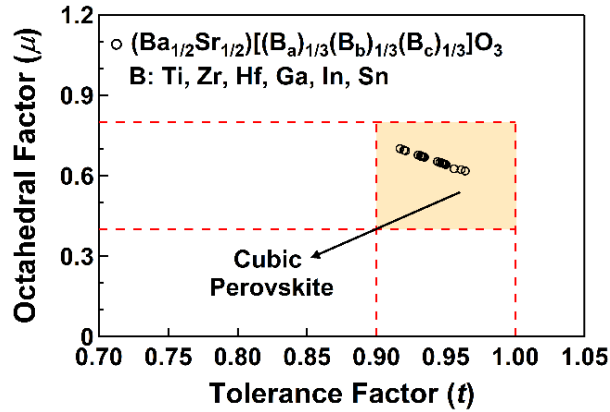

Fig 1. Calculated tolerance factor and octahedral factor for 20 possible compositions with general formula $(Ba_{1/2}Sr_{1/2})[(B_a)_{1/3}(B_b)_{1/3}(B_c)_{1/3}]O_3$ (B: Ti, Zr, Hf, Ga, In, Sn) and desirable range of these two factors for forming high-entropy perovskites with cubic crystal structure.

For experimental synthesis, three compositions were selected to have only $d^0$ B-type cations, only $d^{10}$ B-type cations and $d^0+d^{10}$ B-type cations: $(Ba_{1/2}Sr_{1/2})(Ti_{1/3}Zr_{1/3}Hf_{1/3})O_3$,

$(Ba_{1/2}Sr_{1/2})(Ga_{1/3}In_{1/3}Sn_{1/3})O_{3-\delta}$ and $(Ba_{1/2}Sr_{1/2})(Ti_{1/3}Zr_{1/3}Sn_{1/3})O_3$. These combinations of cationic configurations were selected because an earlier study suggested the significance of $d^0$, $d^{10}$ or $d^0+d^{10}$ on photocatalytic performance [41,42]. Table 1 shows that the configuration entropy and tolerance factor of these three HEPs are greater than $1.5 R$ and within 0.9-1.0, respectively. Additionally, considering the octahedral factor in the range of 0.41 to 0.73 [43] and the valence deviation and ion-radius deviation greater than 0.3 [24], the three materials meet all the necessary conditions to achieve a high-entropy state and form a cubic phase in the crystal structure. It should be noted that, since in $ABX_3$ perovskites, B-site cations have a valence of +4, the use of cations with a lower valence, such as $Ga^{3+}$ and $In^{3+}$, leads to a non-stoichiometric oxygen ratio below 3, as in $(Ba_{1/2}Sr_{1/2})(Ga_{1/3}In_{1/3}Sn_{1/3})O_{3-\delta}$.

Table 1. Theoretical parameters used to design high-entropy perovskites $(Ba_{1/2}Sr_{1/2})(Ti_{1/3}Zr_{1/3}Hf_{1/3})O_3$, $(Ba_{1/2}Sr_{1/2})(Ti_{1/3}Zr_{1/3}Sn_{1/3})O_3$ and $(Ba_{1/2}Sr_{1/2})(Ga_{1/3}In_{1/3}Sn_{1/3})O_{3-\delta}$. $R$ is gas constant.

| Composition | $(Ba_{1/2}Sr_{1/2})(Ti_{1/3}Zr_{1/3}Hf_{1/3})O_3$ | $(Ba_{1/2}Sr_{1/2})(Ti_{1/3}Zr_{1/3}Sn_{1/3})O_3$ | $(Ba_{1/2}Sr_{1/2})(Ga_{1/3}In_{1/3}Sn_{1/3})O_{3-\delta}$ |
|---|---|---|---|
| Entropy configuration, $\Delta S_{config}$ | $1.79R$ | $1.79R$ | $1.79R$ |
| Tolerance factor, $t$ | 0.946 | 0.949 | 0.935 |
| Octahedra factor, $\mu$ | 0.649 | 0.644 | 0.669 |
| Octahedra mismatch, $\Delta\mu$ | 0.061 | 0.016 | 0.058 |
| Valence deviation, $\delta(V)$ | 0.451 | 0.451 | 0.382 |
| Ion-radius deviation, $\delta(r)$ | 0.468 | 0.474 | 0.448 |

## 2.2. Synthesis of catalysts

The different binary oxides, including barium peroxide ($BaO_2$, 99 %), strontium peroxide ($SrO_2$, 98 %), anatase titanium dioxide ($TiO_2$, 98 %), zirconium dioxide ($ZrO_2$, 97 %), hafnium dioxide ($HfO_2$, 98 %), gallium trioxide ($Ga_2O_3$, 99.99 %), indium trioxide ($In_2O_3$, 99.99 %) and tin dioxide ($SnO_2$, 99.9 %), were purchased from Sigma-Aldrich, USA, and Kojundo, Japan. To synthesize HEPs, the oxides were weighed based on the atomic percent ratio of each cation in the proposed structure. They were then poured into a mortar and ground in ethanol for about 30 min to form a mixture. Next, this mixture was compressed into a cylindrical disc ($V = 0.063$ cm$^3$, $r = 0.5$ cm) for HPT processing ($P = 6$ GPa, $N = 3$ turns, $\omega = 1$ rpm, $T = 298$ K). The process of grinding and HPT processing was repeated two times before the mixture was placed in a furnace ($T = 1373$ K, $t = 24$ h). Finally, to ensure the material is fully homogenized at the atomic scale, the sample was treated once more by HPT and calcination.

## 2.3. Characterization of catalysts

Various analytical techniques were employed to investigate the properties of the three HEPs. Among them, X-ray diffraction (XRD) with a Cu Kα radiation source was used to analyse the crystal lattice of the materials. In addition, the characteristics of the A-site and B-site cations in the perovskite structure were also examined through different vibrational modes by Raman spectroscopy (laser lamp with a wavelength of 532 nm). To confirm the microscale and nanoscale features of the HEPs, scanning electron microscopy (SEM) technique at 15 keV was used to collect backscattered electron (BSE) images, along with the transmission electron microscopy (TEM) method at 200 keV to obtain bright-field (BF) images, dark-field (DF) images, selected area electron diffraction (SAED) and high-resolution (HR) images. To determine the elemental homogeneity of the materials, energy-dispersive X-ray spectroscopy (EDS) combined with SEM and scanning-transmission electron microscopy (STEM) was applied. UV-Vis spectroscopy combined with the Kubelka-Munk method was performed to determine the light absorption of the catalyst in the range of 200-800 nm and estimate the bandgap. To investigate the oxidation-reduction state and surface characteristics, X-ray photoelectron spectroscopy (XPS) was used with an Al Kα radiation source. To estimate the valence band top, ultraviolet photoelectron spectroscopy (UPS) was conducted by employing a helium UV light source. The recombination of charge carriers (electrons and holes) was measured using the photoluminescence method with a laser source with a 325 nm wavelength. Finally, to investigate oxygen vacancies in the materials, the electron spin resonance (ESR) technique was conducted at a frequency of 9.4688 GHz to study the behavior of unpaired electrons.

## 2.4. Water splitting experiments

Photocatalytic experiments for $H_2$ generation from water splitting were carried out with 50 mg of HEPs and 27 mL of deionized $H_2O$ in a quartz bottle sealed with a septum cap and purged with argon gas for 10 min. Additionally, 3 mL of methyl alcohol was added as a scavenger and 0.25 mL of 0.01 M $H_2PtCl_6 \cdot 6H_2O$ was added as a source of platinum cocatalyst in some experiments. The experimental system was set up using an xenon lamp (300 W, 1.3 W/cm$^2$) as the radiation source, positioned approximately 10 cm away from the quartz photoreactor. The photoreactor was placed in a cooling tank, and a magnetic stirrer was used to maintain a constant stirring speed of 420 rpm throughout the experiment. The $H_2$ concentration was monitored by directly injecting 500 μL of the gas phase into a gas chromatograph (GC) having a thermal conductivity detector (TCD).

## 3. Results

### 3.1. Catalyst characterization

Fig. 2 shows the XRD profiles to describe the crystal structure of three HEPs. As shown in Fig. 2a, 2c and 2e, the XRD profiles depict the structure transformation process of the initial mixed oxides through five processing steps combining HPT and calcination. The final products are HEPs with an ideal single-phase cubic structure in the space group of Pm$\bar{3}$m. The existence of single-phase perovskites is further confirmed through the corresponding Rietveld refinement analysis, as shown in Fig. 2b, 2d and 2f. Table 2 shows the lattice parameters of perovskites and the $R_{wp}$ (weighted profile residual) and $R_p$ (profile residual) values for their Rietveld analysis, which are below 10 %, indicating good refinements. These experimental structural analyses confirm the validity of the theoretical calculations used for the design of materials.

Fig. 2g and Table 3 describe the Raman spectra of the three HEPs and provide information on the vibrational modes corresponding to the observed peak positions. Comparing Raman spectra achieved for each HEP at three random sites indicates a high uniformity and single-phase structure, as the Raman spectra show almost no significant differences. Through the examination of the spectra of the three different HEPs, the different peak positions correspond to different vibrational modes, thereby providing a clearer understanding of the structure of each HEP. First, the peak position at 144-152 cm$^{-1}$ is assigned to the lattice vibrations [11]. Second, the peak observed at 249-254 cm$^{-1}$ relates to the A-site cations, describing the vibrations of Ba-O or Sr-O bonds in the crystal [44]. Third, the vibration of oxygen atoms in the octahedra ($B_1/B_2/B_3$)$O_6$ is based on a peak at 524 cm$^{-1}$ [45–47] and a dominant peak in the range of 707-718 cm$^{-1}$ [45–48]. Finally, the peak at 400 cm$^{-1}$ is characteristic of oxygen vacancy sites [44], and the peak at 852-862 cm$^{-1}$ is characteristic of point defects, including vacancies [45]. The peak at 400 cm$^{-1}$ is observed only in (Ba$_{1/2}$Sr$_{1/2}$)(Ga$_{1/3}$In$_{1/3}$Sn$_{1/3}$)O$_{3-\delta}$, suggesting a very high oxygen vacancy concentration in this material, compared to two other HEPs. The high fraction of oxygen vacancies in (Ba$_{1/2}$Sr$_{1/2}$)(Ga$_{1/3}$In$_{1/3}$Sn$_{1/3}$)O$_{3-\delta}$ can be explained by the charge deficiency of the three constituent B-site cations (Ga$^{3+}$, In$^{3+}$, Sn$^{4+}$) in the A$^{2+}$B$^{4+}$X$_3^{2-}$ structure, leading to a non-stoichiometric oxygen ratio as well as the creation of oxygen vacancies (no charge deficiency in B sites for the other two HEPs). These Raman results further confirm the successful synthesis of single-phase cubic HEP materials. Fig. 2h depicts the three-dimensional constructed crystal structure for these materials based on structural analyses by XRD and Raman spectroscopy.

As described in Fig. 3, the HEP samples were determined for their composition through two methods: SEM-EDS and STEM-EDS. Fig. 3a, 3c and 3e show a homogeneous distribution at the micrometer scale, while Fig. 3b, 3d and 3f demonstrate homogeneity at the nanometer scale for (a, b) $(Ba_{1/2}Sr_{1/2})(Ti_{1/3}Zr_{1/3}Hf_{1/3})O_3$, (c, d) $(Ba_{1/2}Sr_{1/2})(Ti_{1/3}Zr_{1/3}Sn_{1/3})O_3$ and (e, f) $(Ba_{1/2}Sr_{1/2})(Ga_{1/3}In_{1/3}Sn_{1/3})O_{3-\delta}$. Additionally, the elemental composition percentages are presented in Table 4, indicating an acceptable consistency with the nominal atomic composition in the compounds, i.e. 10.0 at% for each A-type cation, 6.7 at% for each B-type cation, and 60 at% for oxygen.

Fig. 4 describes the microscopic characteristics of the three HEPs. The TEM (a, b, c) BF and (d, e, f) DF images indicate the presence of a large number of nanograins. In addition, the SAED patterns in Fig. 4g, 4h and 4i exhibit concentric rings, resulting from the random orientation of nanograins with the cubic structure, where each ring represents a specific crystal plane in the HEPs. A more detailed observation of each material is also provided through HR images and fast Fourier transform (FFT) analyses in Fig. 4j, 4k and 4l, confirming that all three HEPs have the cubic phase in their microstructure. These results further suggest the formation of single-phase HEPs.

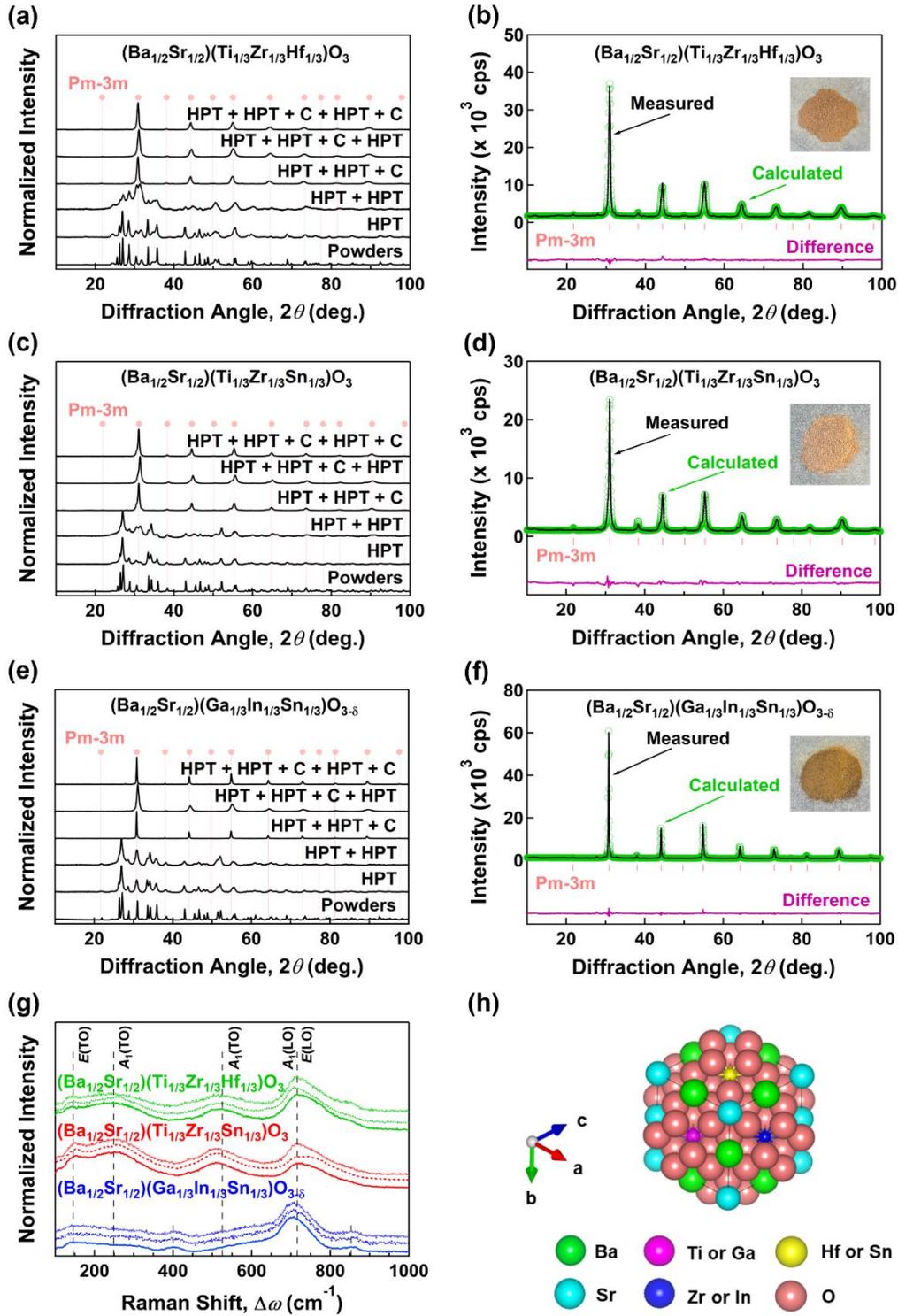

Fig. 2. Synthesis process for forming single-phase cubic structure in high-entropy perovskites. (a, c, e) XRD patterns at five synthesis stages performed by HPT ($N$ = 3 turns, $\omega$ = 1 rpm, $T$ = 297 K) and calcination ($T$ = 1373 K, $t$ = 24 h) and (b, d, f) XRD profile and corresponding Rietveld analysis of (a, b) $(Ba_{1/2}Sr_{1/2})(Ti_{1/3}Zr_{1/3}Hf_{1/3})O_3$, (c, d) $(Ba_{1/2}Sr_{1/2})(Ti_{1/3}Zr_{1/3}Sn_{1/3})O_3$ and (e, f) $(Ba_{1/2}Sr_{1/2})(Ga_{1/3}In_{1/3}Sn_{1/3})O_{3-\delta}$. (g) Raman spectra of perovskites examined at three

different sites. (h) Model crystal structure of cubic perovskites. Inset in (b, d, f): appearance of samples.

Table 2. Experimental lattice parameters and Rietveld refinement factors $R_{wp}$ (weighted profile residual) and $R_p$ (profile residual) for single-phase cubic high-entropy perovskites $(Ba_{1/2}Sr_{1/2})(Ti_{1/3}Zr_{1/3}Hf_{1/3})O_3$, $(Ba_{1/2}Sr_{1/2})(Ti_{1/3}Zr_{1/3}Sn_{1/3})O_3$ and $(Ba_{1/2}Sr_{1/2})(Ga_{1/3}In_{1/3}Sn_{1/3})O_{3-\delta}$.

| Composition | $(Ba_{1/2}Sr_{1/2})(Ti_{1/3}Zr_{1/3}Hf_{1/3})O_3$ | $(Ba_{1/2}Sr_{1/2})(Ti_{1/3}Zr_{1/3}Sn_{1/3})O_3$ | $(Ba_{1/2}Sr_{1/2})(Ga_{1/3}In_{1/3}Sn_{1/3})O_{3-\delta}$ |
|---|---|---|---|
| Unit-cell parameters | $a = b = c = 4.08982(4)$ Å<br>$V = 68.41$ Å$^3$ | $a = b = c = 4.06619(5)$ Å<br>$V = 67.23$ Å$^3$ | $a = b = c = 4.09784(2)$ Å<br>$V = 68.81$ Å$^3$ |
| Rietveld analysis | $R_{wp} = 4.6$ %<br>$R_p = 3.6$ % | $R_{wp} = 7.0$ %<br>$R_p = 5.1$ % | $R_{wp} = 8.4$ %<br>$R_p = 6.7$ % |

Table 3. Raman peak positions and corresponding vibrational modes for high-entropy perovskites $(Ba_{1/2}Sr_{1/2})(Ti_{1/3}Zr_{1/3}Hf_{1/3})O_3$, $(Ba_{1/2}Sr_{1/2})(Ti_{1/3}Zr_{1/3}Sn_{1/3})O_3$ and $(Ba_{1/2}Sr_{1/2})(Ga_{1/3}In_{1/3}Sn_{1/3})O_{3-\delta}$.

| Raman shift (cm$^{-1}$) | | | | | |
|---|---|---|---|---|---|
| $(Ba_{1/2}Sr_{1/2})(Ti_{1/3}Zr_{1/3}Hf_{1/3})O_3$ | $(Ba_{1/2}Sr_{1/2})(Ti_{1/3}Zr_{1/3}Sn_{1/3})O_3$ | $(Ba_{1/2}Sr_{1/2})(Ga_{1/3}In_{1/3}Sn_{1/3})O_{3-\delta}$ | Vibrational mode | Symmetry | Ref. |
| 145.8 | 151.6 | 143.9 | Lattice | $E$ (TO) | [11] |
| 249.0 | 253.8 | - | Ba/Sr-O vibrations | $A_1$ (TO) | [44] |
| - | - | 399.8 | Oxygen vacancy | - | [44] |
| 525.5 | 523.6 | - | Vibration of oxygen atoms in (Ti/Zr/Hf)O$_6$, (Ti/Zr/Sn)O$_6$ and (Ga/In/Sn)O$_6$ octahedra | $A_1$ (TO) | [45–47] |
| 716.3 | 718.1 | 707.3 | | $A_1$ (LO), $E$ (LO) | [45–48] |
| 853.1 | - | 862.0 | Point defects such as vacancies | - | [45] |

Table 4. Elemental composition of high-entropy perovskites $(Ba_{1/2}Sr_{1/2})(Ti_{1/3}Zr_{1/3}Hf_{1/3})O_3$, $(Ba_{1/2}Sr_{1/2})(Ti_{1/3}Zr_{1/3}Sn_{1/3})O_3$ and $(Ba_{1/2}Sr_{1/2})(Ga_{1/3}In_{1/3}Sn_{1/3})O_{3-\delta}$ measured by SEM-EDS and STEM-EDS.

| High-entropy perovskite | Technique | Alkaline earth metal (at%) | | $d^0$ metal (at%) | | | $d^{10}$ metal (at%) | | | Anion (at%) |
|---|---|---|---|---|---|---|---|---|---|---|
| | | Ba | Sr | Ti | Zr | Hf | Ga | In | Sn | O |
| $(Ba_{1/2}Sr_{1/2})(Ti_{1/3}Zr_{1/3}Hf_{1/3})O_3$ | SEM-EDS | 9.7±1.7 | 8.5±0.5 | 6.4±0.6 | 5.3±1.2 | 5.7±0.8 | - | - | - | 64.4±4.2 |
| | STEM-EDS | 7.7 | 6.6 | 6.3 | 5.9 | 4.1 | - | - | - | 69.4 |
| $(Ba_{1/2}Sr_{1/2})(Ti_{1/3}Zr_{1/3}Sn_{1/3})O_3$ | SEM-EDS | 9.5±1.4 | 8.1±1.1 | 6.1±2.8 | 4.2±1.6 | - | - | - | 6.5±2.1 | 65.7±4.2 |
| | STEM-EDS | 8.4 | 8.0 | 7.5 | 5.2 | | - | - | 6.0 | 64.9 |
| $(Ba_{1/2}Sr_{1/2})(Ga_{1/3}In_{1/3}Sn_{1/3})O_{3-\delta}$ | SEM-EDS | 10.7±1.9 | 9.1±0.9 | - | - | - | 7.4±3.4 | 6.2±0.8 | 7.5±1.1 | 59.2±6.8 |
| | STEM-EDS | 10.0 | 14.0 | - | - | - | 8.3 | 7.3 | 5.5 | 54.9 |

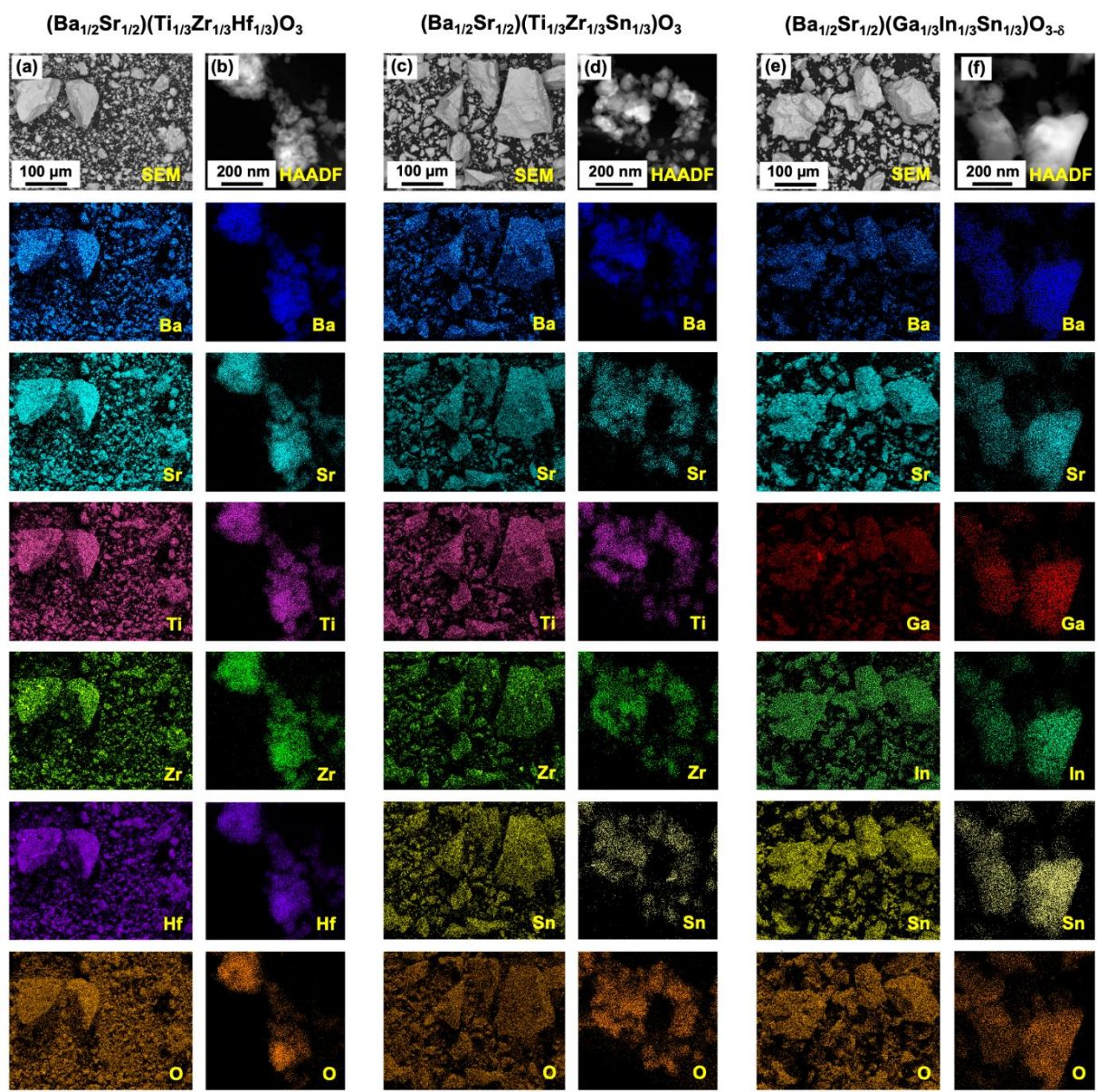

Fig. 3. Uniform distribution of elements in high-entropy perovskite photocatalysts at micrometer and nanometer scales. (a, c, e) SEM images with EDS mappings and (b, d, f) HAADF images taken by STEM and related EDS mappings for (a, b) $(Ba_{1/2}Sr_{1/2})(Ti_{1/3}Zr_{1/3}Hf_{1/3})O_3$, (c, d) $(Ba_{1/2}Sr_{1/2})(Ti_{1/3}Zr_{1/3}Sn_{1/3})O_3$ and (e, f) $(Ba_{1/2}Sr_{1/2})(Ga_{1/3}In_{1/3}Sn_{1/3})O_{3-\delta}$.

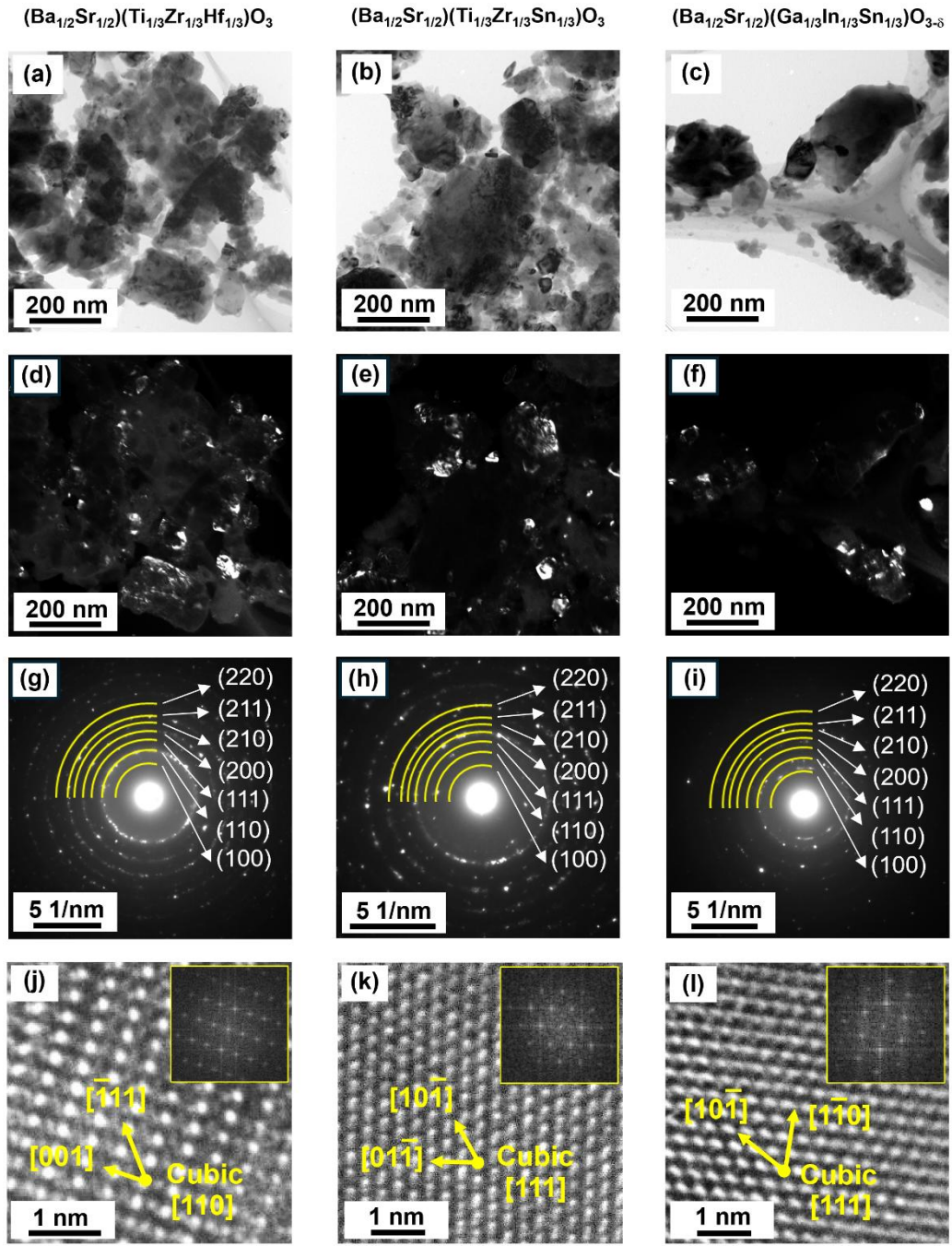

Fig. 4. Nanograined single-phase cubic structure of high-entropy perovskites. (a, b, c) TEM BF micrographs, (d, e, f) DF micrographs, (g, h, i) SAED patterns, (j, k, l) HR images with FFT diffractograms for (a, d, g, j) $(Ba_{1/2}Sr_{1/2})(Ti_{1/3}Zr_{1/3}Hf_{1/3})O_3$, (b, e, h, k) $(Ba_{1/2}Sr_{1/2})(Ti_{1/3}Zr_{1/3}Sn_{1/3})O_3$ and (c, f, i, l) $(Ba_{1/2}Sr_{1/2})(Ga_{1/3}In_{1/3}Sn_{1/3})O_{3-\delta}$.

The oxidation-reduction states of the elements in the three HEPs are shown in Fig. 5 using XPS. The spectra indicate the presence of peaks at 779.1 eV for Ba $3d^{5/2}$, 132.8-133.4 eV for Sr $3d^{5/2}$, 457.9 eV for Ti $2p^{3/2}$, 180.8 eV for Zr $3d^{5/2}$, 16.0 eV for Hf $4f^{7/2}$, 1116.8 eV for Ga $2p^{3/2}$, 444.0 eV for In $3d^{5/2}$ and 485.8 eV for Sn $3d^{5/2}$. These results confirm the complete oxidation states of the cations in the HEPs. In addition, when comparing the HEPs, a shift to lower energy is observed for the Sr $3d^{5/2}$ peak of $(Ba_{1/2}Sr_{1/2})(Ga_{1/3}In_{1/3}Sn_{1/3})O_{3-\delta}$, compared to the other two HEPs. In contrast, the O 1s peak of $(Ba_{1/2}Sr_{1/2})(Ga_{1/3}In_{1/3}Sn_{1/3})O_{3-\delta}$ at 530.2 eV appears at a higher energy level than 529.4 eV for $(Ba_{1/2}Sr_{1/2})(Ti_{1/3}Zr_{1/3}Hf_{1/3})O_3$ and $(Ba_{1/2}Sr_{1/2})(Ti_{1/3}Zr_{1/3}Sn_{1/3})O_3$. This O 1s peak shift to higher energies indicates the oxygen-deficient structure of this catalyst. The peak deconvolution also shows the presence of a peak at 531.8 eV, associated with the adsorption of OH groups and oxygen vacancy sites [49] in the HEPs. These results demonstrate that oxidation states of the HEPs fit the $A^{2+}B^{4+}X_3^{2-}$ formula, but some vacancies exist.

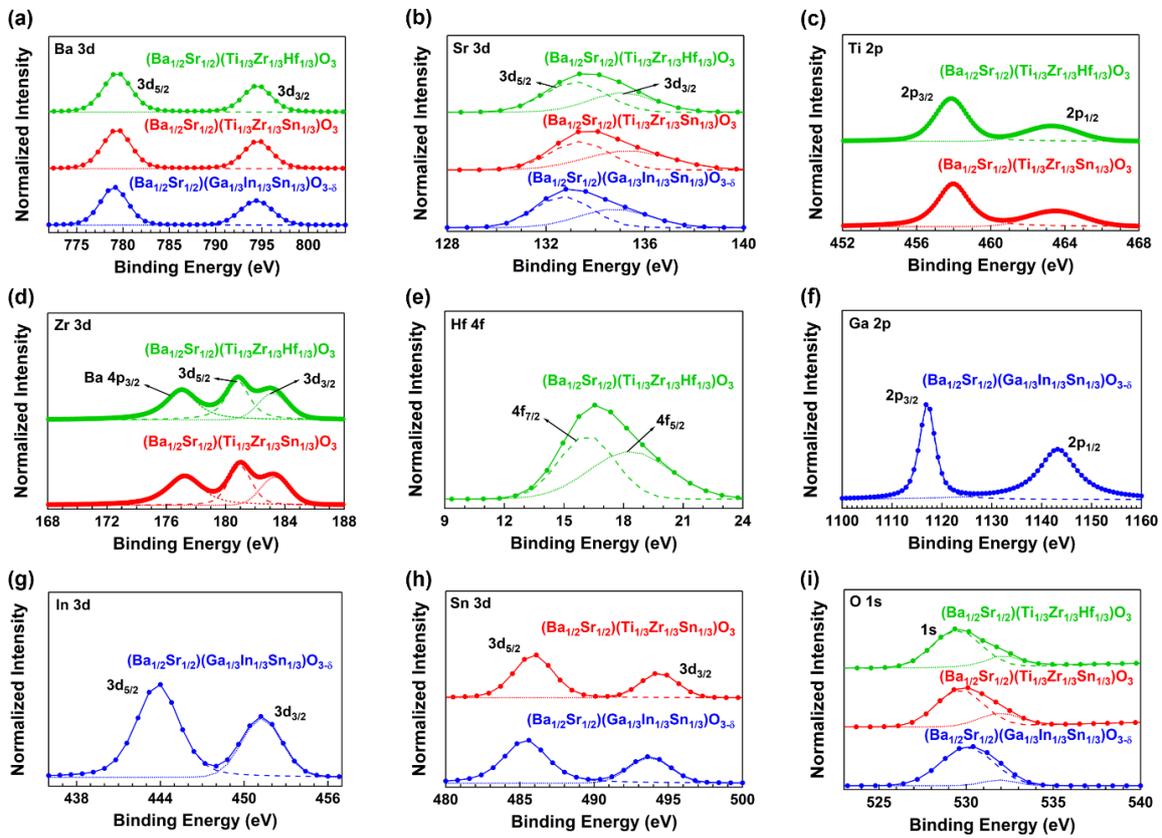

Fig. 5. Valence states of the cations and oxygen in three high-entropy perovskites. XPS spectra and coresponding peak deconvolution of (a) Ba 3d, (b) Sr 3d, (c) Ti 2p, (d) Zr 3d, (e) Hf 4f, (f) Ga 2p, (g) In 3d, (h) Sn 3d and (i) O 1s for (a-e, i) $(Ba_{1/2}Sr_{1/2})(Ti_{1/3}Zr_{1/3}Hf_{1/3})O_3$, (a-d, h, i) $(Ba_{1/2}Sr_{1/2})(Ti_{1/3}Zr_{1/3}Sn_{1/3})O_3$ and (a, b, f-i) $(Ba_{1/2}Sr_{1/2})(Ga_{1/3}In_{1/3}Sn_{1/3})O_{3-\delta}$.

The optical properties, band structure, oxygen vacancy and electron-hole recombination rates in HEPs were determined by analytical methods such as UV-Vis absorbance, UPS, ESR and Photoluminescence, as shown in Fig. 6. Fig. 6a shows the UV-Vis absorbance spectra of all three HEPs, each displaying high light absorbance in the UV light region. In addition, some visible-light absorbance is observed in the three HEPs. Fig. 6b determines the bandgap based on the Kubelka-Munk method, with values of 3.1 eV, 3.3 eV and 2.8 eV for $(Ba_{1/2}Sr_{1/2})(Ti_{1/3}Zr_{1/3}Hf_{1/3})O_3$, $(Ba_{1/2}Sr_{1/2})(Ti_{1/3}Zr_{1/3}Sn_{1/3})O_3$ and $(Ba_{1/2}Sr_{1/2})(Ga_{1/3}In_{1/3}Sn_{1/3})O_{3-\delta}$, respectively. Fig. 6c describes the valence band maximum (VBM) of the materials based on bias-corrected UPS spectra, in which the VBM position is estimated based on a tail at low energy levels, while the secondary electron cut-off (SECO) is estimated from the position of the other tail at high energy levels. The measured VBM and SECO values are $3.5 \pm 0.1$ eV versus Fermi level ($E_F$) and $17.3 \pm 0.1$ eV versus $E_F$ for $(Ba_{1/2}Sr_{1/2})(Ti_{1/3}Zr_{1/3}Hf_{1/3})O_3$; $2.9 \pm 0.1$ eV versus $E_F$ and $16.9 \pm 0.1$ eV versus $E_F$ for $(Ba_{1/2}Sr_{1/2})(Ti_{1/3}Zr_{1/3}Sn_{1/3})O_3$; $3.8 \pm 0.1$ eV versus $E_F$ and $17.7 \pm 0.1$ eV versus $E_F$ for $(Ba_{1/2}Sr_{1/2})(Ga_{1/3}In_{1/3}Sn_{1/3})O_{3-\delta}$, respectively. The SECO value is used to calculate the work function (i.e. Fermi level versus vacuum, $\phi$) as the difference between SECO and the He I radiation energy ($h\upsilon$ = 21.2 eV, h: Planck's constant, $\upsilon$: photon frequency).

$$\phi = h\upsilon - E_{SECO} \qquad (6)$$

From the work function and the VBM versus the Fermi level, the VBM versus vacuum can be calculated.

$$E_{VBM\ vs.\ vacuum} = -(E_{VBM\ vs.\ Fermi\ level} + \phi) \qquad (7)$$

$$E_{VBM\ vs.\ vacuum} = -(E_{VBM\ vs.\ Fermi\ level} + h\upsilon - E_{SECO}) \qquad (8)$$

Eq. 8 is not affected by biasing during UPS, because the effects of bias on VBM versus the Fermi level and SECO cancel out each other. The VBM versus vacuum is achieved $-7.4 \pm 0.2$ eV, $-7.2 \pm 0.2$ eV and $-7.3 \pm 0.2$ eV for $(Ba_{1/2}Sr_{1/2})(Ti_{1/3}Zr_{1/3}Hf_{1/3})O_3$, $(Ba_{1/2}Sr_{1/2})(Ti_{1/3}Zr_{1/3}Sn_{1/3})O_3$ and $(Ba_{1/2}Sr_{1/2})(Ga_{1/3}In_{1/3}Sn_{1/3})O_{3-\delta}$, respectively. Based on the calculated bandgap and VBM values, an estimated band structure of the catalysts is illustrated in Fig. 6d, in which the conduction band minimum (CBM) was calculated as the difference between the VBM and corresponding bandgaps for each HEP. It is noticed that all three materials have band structures with their VBM lower than the $O_2/H_2O$ oxidation potential, while their CBM is higher than the $H^+/H_2$ reduction potential.

The ESR method was used to confirm the presence of vacancy defects. As observed in Fig. 6e, all three ESR spectra exhibit dual peaks with a turning point at 2.008, confirming the presence of oxygen-based defects with unpaired electrons [50] in the HEPs. These EPR results, which are also consistent with the O 1s XPS analysis and somehow different from the Raman spectroscopy data, suggest that the oxygen vacancies are present in three HEPS, but their concentration is higher in $(Ba_{1/2}Sr_{1/2})(Ga_{1/3}In_{1/3}Sn_{1/3})O_{3-\delta}$ because of the charge deficiency of the constituent B-site cations. Fig. 6f illustrates the photoluminescence spectra, showing a peak around 580 nm for all three materials with close intensities, but the intensity for $(Ba_{1/2}Sr_{1/2})(Ti_{1/3}Zr_{1/3}Sn_{1/3})O_3$ is somewhat higher than the other two. These results demonstrate the suitability of HEP materials for photocatalytic reactions, such as water splitting to produce $H_2$.

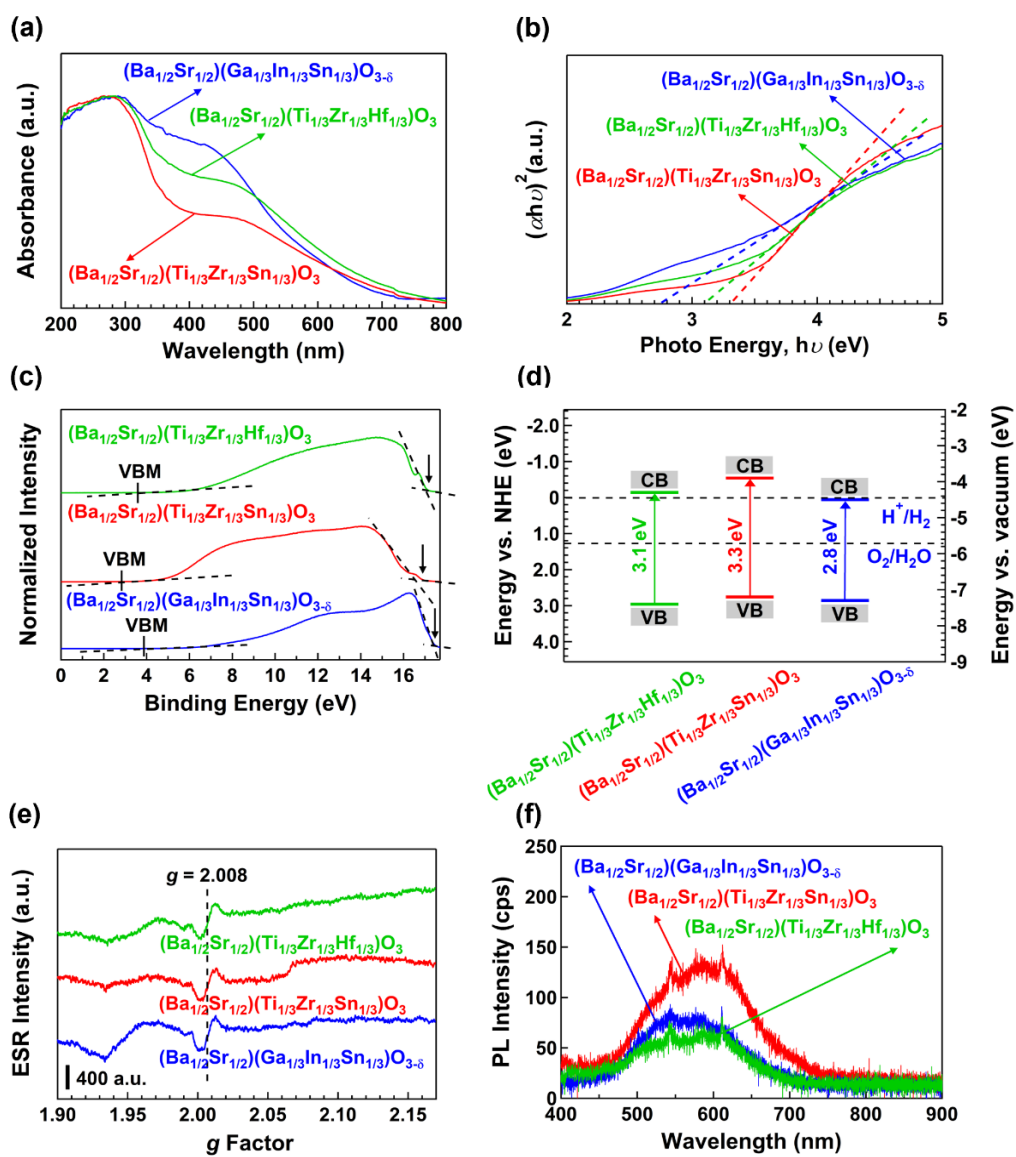

Fig. 6. Optical properties and suitable band structure of high-entropy perovskites for water splitting. (a) UV-Vis absorbance spectra, (b) Kulbelka-Munk method for direct bandgap calculation, (c) bias-corrected UPS spectra (He I, bias -2 V) for VBM determination, (d) band structure, (e) ESR spectra and (f) PL spectra for $(Ba_{1/2}Sr_{1/2})(Ti_{1/3}Zr_{1/3}Hf_{1/3})O_3$, $(Ba_{1/2}Sr_{1/2})(Ti_{1/3}Zr_{1/3}Sn_{1/3})O_3$ and $(Ba_{1/2}Sr_{1/2})(Ga_{1/3}In_{1/3}Sn_{1/3})O_{3-\delta}$.

## 3.2. Photocatalytic activity

Fig. 7 describes the formation of $H_2$ through water-splitting using (a) $(Ba_{1/2}Sr_{1/2})(Ti_{1/3}Zr_{1/3}Hf_{1/3})O_3$, (b) $(Ba_{1/2}Sr_{1/2})(Ti_{1/3}Zr_{1/3}Sn_{1/3})O_3$ and (c) $(Ba_{1/2}Sr_{1/2})(Ga_{1/3}In_{1/3}Sn_{1/3})O_{3-\delta}$ under 3 h of light irradiation. The experiment for each HEP was conducted under three different experimental conditions. In Experiment 1, only HEP and deionized water were used. Experiment 2 was similar to Experiment 1 but included methanol as a scavenger. Experiment 3 was similar to Experiment 2 but with the addition of $H_2PtCl_6 \cdot 6H_2O$ as a source of platinum cocatalyst. The HEPs produce only a very small amount of $H_2$, ranging from 0.007 to 0.02 mmol/g after 3 h under the condition of using only the catalyst and water. After the addition of methanol, the efficiency of $H_2$ generation improves significantly, reaching 0.9-1.1 mmol/g after 3 h. However, this efficiency does not increase significantly with the addition of the cocatalyst, and the value remains almost unchanged. No changes in the photocatalytic $H_2$ production by the addition of $H_2PtCl_6 \cdot 6H_2O$ are not due to the absence of platinum on the surface, because the color of all samples changed to black by photodeposition of platinum. It should be noted that $H_2PtCl_6 \cdot 6H_2O$ shows very high activity for photodeposition of platinum during UV irradiation, and its effectiveness has also been proven for Ba- and Sr-based perovskites [51–53]. The $H_2$ concentration formed from the three different HEPs showed similar values. To evaluate the reusability of catalysts, three consecutive cycles were conducted, as illustrated in Fig. 7d. The $H_2$ production rate did not show any decrease after repeating cycles. Fig. 8 examines the structural stability of catalysts through conducting (a) XRD analysis and (b-e) SEM analysis (b, d, f) before and (c, e, g) after the catalysis, showing that the catalyst structure, morphology and surface are maintained and not affected by the harsh irradiation conditions. These findings highlight that the HEPs have high potential for $H_2$ production without the need for a cocatalyst.

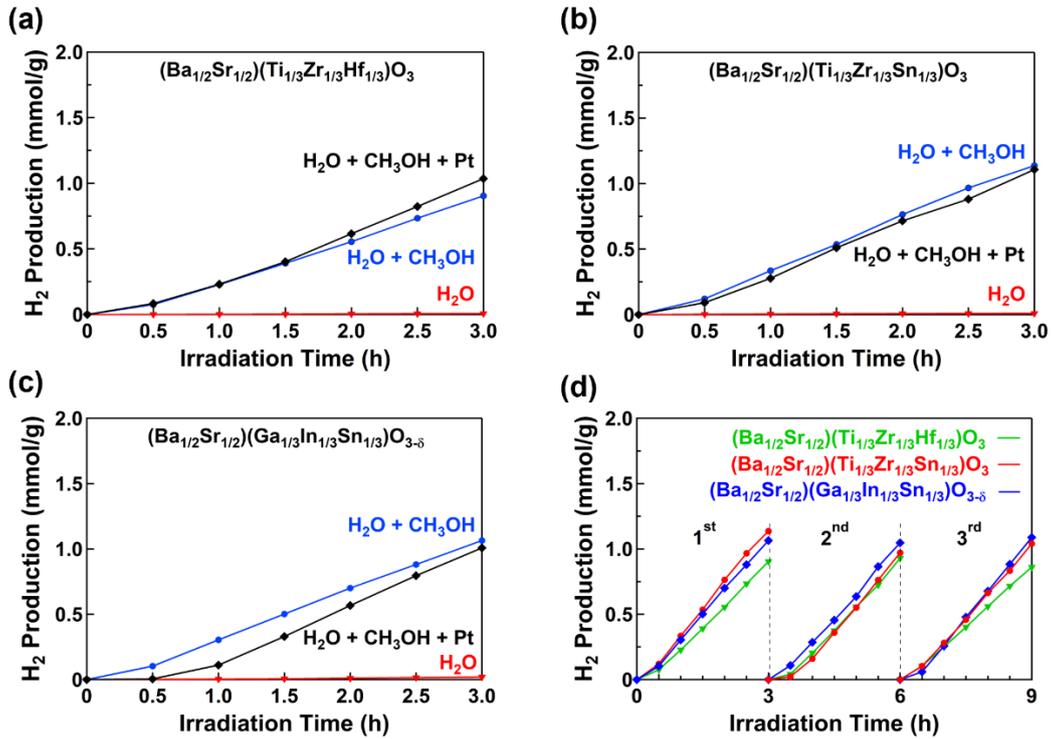

Fig. 7. Photocatalytic activity of three high-entropy perovskites for cocatalyst-free water splitting. (a-c) $H_2$ evolution from water under three conditions of (i) without scavenger and cocatalyst, (ii) with scavenger and without cocatalyst, and (iii) with scavenger and cocatalyst for (a) $(Ba_{1/2}Sr_{1/2})(Ti_{1/3}Zr_{1/3}Hf_{1/3})O_3$, (b) $(Ba_{1/2}Sr_{1/2})(Ti_{1/3}Zr_{1/3}Sn_{1/3})O_3$ and (c) $(Ba_{1/2}Sr_{1/2})(Ga_{1/3}In_{1/3}Sn_{1/3})O_{3-\delta}$. (d) Reusability of perovskites for $H_2$ production from water splitting with scavenger and without cocatalyst in three consecutive cycles.

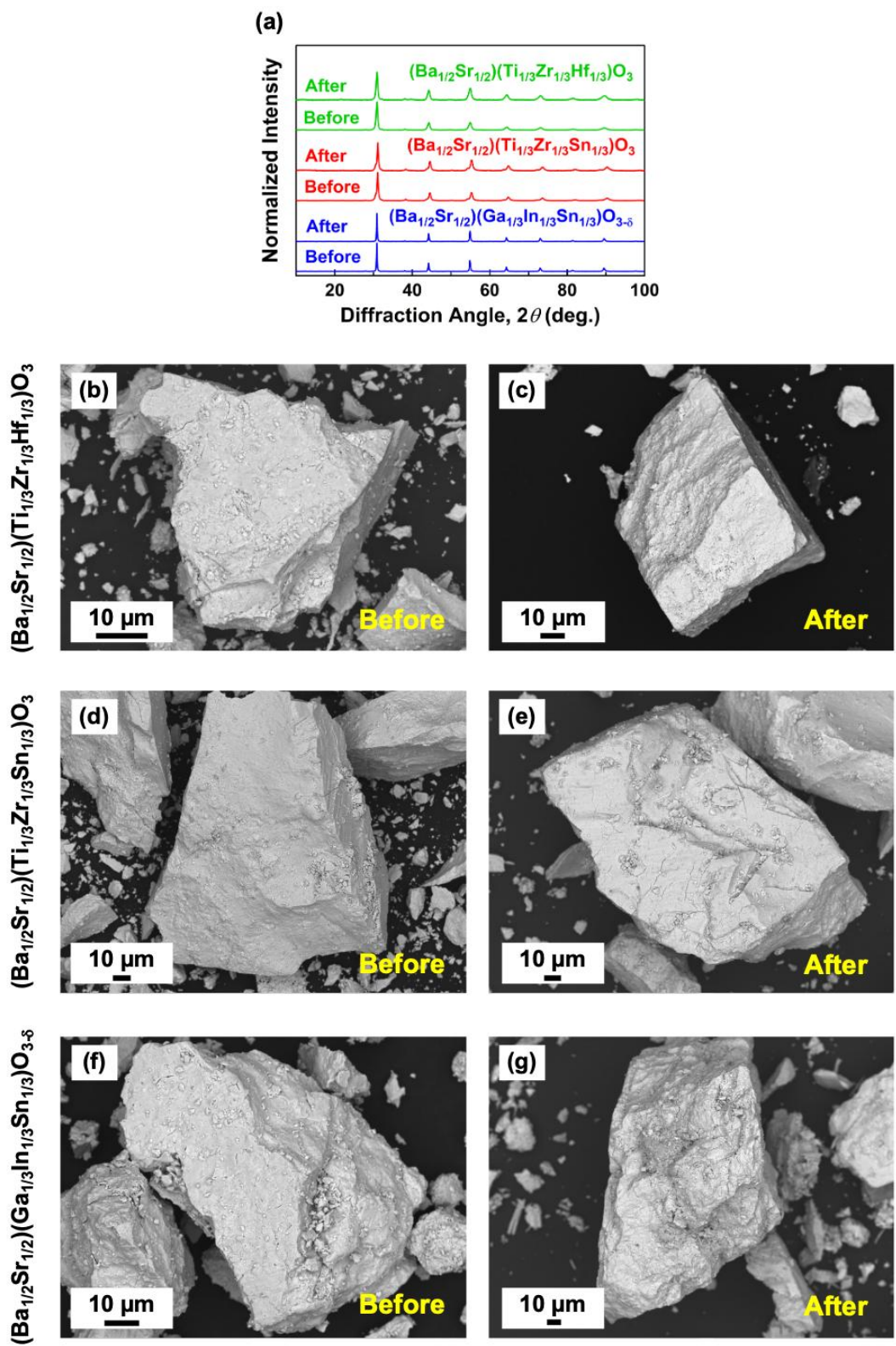

Fig. 8. Stability of high-entropy perovskites for hydrogen evolution. (a) XRD spectra before and after 3 cycles of water splitting. (b-g) SEM micrographs of (a, c) (Ba$_{1/2}$Sr$_{1/2}$)(Ti$_{1/3}$Zr$_{1/3}$Hf$_{1/3}$)O$_3$, (d, e) (Ba$_{1/2}$Sr$_{1/2}$)(Ti$_{1/3}$Zr$_{1/3}$Sn$_{1/3}$)O$_3$ and (f, g) (Ba$_{1/2}$Sr$_{1/2}$)(Ga$_{1/3}$In$_{1/3}$Sn$_{1/3}$)O$_{3-\delta}$ (b, d, f) before and (c, e, g) after photocatalytic water splitting.

## 4. Discussion

To better understand the behavior of HEPs in the water-splitting process without using cocatalysts, this section discusses two issues. The first issue is why HEPs do not require a co-catalyst in the photocatalytic water splitting process. The second issue is why the catalytic performance in $H_2$ formation does not differ significantly despite the substitution of B-site positions in HEPs.

To address the first issue, we need to consider the role of cocatalysts. In the water-splitting process, cocatalysts play a significant role in reducing the activation energy for the reduction of protons at the catalytic sites [17]. Therefore, for HEPs, the highly active proton reduction sites on the surface should contribute to enhancing efficiency, like the use of cocatalysts. In an earlier study on $LaFeO_3$ perovskite, it was also shown that nano-crystalline perovskite is capable of producing $H_2$ through water splitting without platinum, with an efficiency of 75% compared to the case with platinum [54]. Another study also demonstrated that $LiTaO_3$ perovskite exhibits similar effectiveness in photocatalytic water splitting without a cocatalyst, based on the highly negative potential of excited electrons in the CBM region [6]. It was shown that the oxygen p-band center shows a strong correlation with photocatalytic activity [55]. The oxygens that bond to B-site cations in the $[BO_6]$ octahedral units can pump electrons for the reduction of protons on B-site cations and oxygen vacancies [56]. This structural feature creates a favorable kinetic condition for $H_2$ production, allowing for lowering the activation energy for electron transfer in the catalyst even without the need for a cocatalyst [56].

Regarding the second issue, all three HEPs have the same lattice structure, a similar composition at the A site, and show equivalent performance in $H_2$ production despite changes in the composition at the B site. Barium and strontium occupy the A-site positions, playing a role in stabilizing the lattice structure [57]. Additionally, these two cations tend to combine with trivalent and tetravalent ions to ensure charge balance as well as geometric compatibility within the HEP structure [57]. The three HEPs also exhibit insignificant differences in optical properties, such as light absorbance and photoluminescence. It is well known from the literature that the oxygen 2p orbitals (px, py, pz) generate primarily the valence band [58,59]. It is also known that for the conduction band, the main influence comes from the $[BO_6]$ octahedral clusters, formed by the hybridization of the oxygen orbitals and the valence orbitals of B-site cations (ndxz, ndxy, ndyz, $ndz^2$ and $ndx^2$-$y^2$) [58,60,61]. The significant contribution of oxygen to both valence and conduction bands can be inferred from the band structure of the three HEPs in this study, which exhibits similarity in the VBM and only minor differences in the CBM (due to changes in the [BO$_6$] octahedral clusters by adding d$^0$ or d$^{10}$ cations). A previous density functional theory (DFT) computational study on Ba-Sr-Ti-based perovskite materials with and without tin addition at the B sites showed that the electron density distribution remains unchanged for Ba-O and Sr-O bonds, as well as along the Ti-O-Ti or Sn-O-Ti linkage paths, in the two cubic symmetric models of Ba$_x$Sr$_{1-x}$TiO$_3$ and Ba$_x$Sr$_{1-x}$TiSnO$_3$ [58]. Furthermore, a study on the correlation between catalytic performance of perovskites and their electronic structure indicated that perovskites with the highest O-band centers are those connected to B-site cations located in the left or right side of the periodic table (titanium in SrTiO$_3$ or copper in LaCuO$_3$) [55]. These studies can explain the similarity in H$_2$ generation performance observed in three HEPs, in which the oxygen bonds with B-site cations ensure comparable charge characteristics, regardless of whether the B-site cations as active sites for H$_2$ production are occupied by d$^0$ or d$^{10}$ cations.

To evaluate the performance of HEPs, Table 5 compares H$_2$ production from photocatalytic water splitting using different types of perovskites [6,54,62-67]. Different conditions in the reaction setup, including the concentration of catalysts and their surface area, can affect the H$_2$ yield in different studies, so caution is needed in this comparison. Table 5 suggests that HEPs have a moderate activity compared with other types of perovskites. However, this activity should be negatively affected by large particle sizes and the small surface area of the HEPs (Fig. 8) because they were synthesized by the HPT method followed by annealing, which always generates large particle sizes [14,68]. Employing chemical synthesis methods to produce nanoparticles can clarify the real activity of HEPs compared to conventional perovskites. Further studies should also consider measuring the apparent quantum efficiency to quantify the photocatalytic activity of HEPs.

In summary, this study introduces the design criteria to generate single-phase HEPs as cocatalyst-free perovskites, while finding the criteria for developing dual-phase or mixed-phase perovskites can be an interesting topic for future studies. DFT calculations can also confirm the active catalytic sites, and charge separation and transfer mechanisms in HEPs with similar A-site cations. Experimental studies on the effect of A-site element variations, such as the ones attempted in conventional and low-entropy oxides, can also further clarify the photocatalytic behavior of HEPs. Moreover, extension of the use of HEPs to other photocatalytic reactions reported for high-entropy photocatalysts, such as methanation [11], ammonia production [69], plastic degradation [14], antibiotic degradation [70] and oxygen evolution [71] should be further explored.

Table 5. Summary of photocatalytic hydrogen production by using different types of perovskites reported in literature in comparison with high-entropy perovskites.

| Material | Type | Catalytic Conditions | Hydrogen Production | Reference |
|---|---|---|---|---|
| $(Ba_{1/2}Sr_{1/2})(Ti_{1/3}Zr_{1/3}Hf_{1/3})O_3$ | High-Entropy Perovskites | - Solution: Water<br>- Lamp: Xenon (18 kW/m$^2$) | 0.30 mmol/g.h | This Study |
| $(Ba_{1/2}Sr_{1/2})(Ti_{1/3}Zr_{1/3}Sn_{1/3})O_3$ | | | 0.38 mmol/g.h | |
| $(Ba_{1/2}Sr_{1/2})(Ga_{1/3}In_{1/3}Sn_{1/3})O_{3-\delta}$ | | | 0.35 mmol/g.h | |
| $LiTaO_3$ | Tantalate Perovskite | - Solution: Water<br>- Lamp: Xenon | 0.06 mmol/g.h | [62] |
| $CsTaO_3$ | Tantalate Perovskite | - Solution: Water<br>- Lamp: Xenon | 0.06 mmol/g.h | [62] |
| $LiTaO_3$ | Tantalate Perovskite | - Solution: Water<br>- Lamp: 450 W Mercury | 3.00 mmol/g.h | [6] |
| $LaFeO_3$ | Perovskites | - Solution: Water<br>- Lamp: Two 200 W Tungsten | 2.50 mmol/g.h | [54] |
| $La_4Ti_3O_{12}$ | (111) Plane-Type Layered Perovskite | - Solution: water<br>- Lamp: 400 W Mercury | 0.18 mmol/g.h | [63] |
| $CaLa_4Ti_4O_{15}$ | | | 0.06 mmol/g.h | |
| $SrLa_4Ti_4O_{15}$ | | | 0.04 mmol/g.h | |
| $BaLa_4Ti_4O_{15}$ | | | 0.10 mmol/g.h | |
| $Ba_3LaNb_3O_{12}$ | | | 0.04 mmol/g.h | |
| $Sr_5Nb_4O_{15}$ | | | 0.06 mmol/g.h | |
| $Ba_5Nb_4O_{15}$ | | | 0.06 mmol/g.h | |
| $SrTiO_3/TiO_2$-Pt | Perovskites with Heterojunction | - Solution: Water<br>- Lamp: 300 W Xenon | 0.21 mmol/g.h | [64] |
| $BaZr_{0.7}Sn_{0.3}O_3$ | Low-entropy perovskites | - Solution: Water<br>- Lamp: 400 W Mercury | 0.69 mmol/g.h | [65] |
| $Na_{0.95}K_{0.05}TaO_3$ | Low-entropy perovskites | - Solution: water<br>- Lamp: 400 W Mercury | 1.00 mmol/g.h | [66] |
| Methylammonium-Pb$(I_{1-x}Br_x)_3$ | Halide perovskites | - Solution: Hydrohalic acid<br>- Lamp: 300 W Xenon with 420 nm Cut-off Filter (300 mW/cm$^2$) | 1.47 mmol/g.h | [67] |

## 5. Conclusions

In this study, three new high-entropy perovskites (HEPs) were successfully synthesized with barium and strontium at the A site and different elements at the B-site, which include the $d^0$, $d^{10}$ and $d^0+d^{10}$ groups. The three HEPs, $(Ba_{1/2}Sr_{1/2})(Ti_{1/3}Zr_{1/3}Hf_{1/3})O_3$, $(Ba_{1/2}Sr_{1/2})(Ga_{1/3}In_{1/3}Sn_{1/3})O_{3-\delta}$ and $(Ba_{1/2}Sr_{1/2})(Ti_{1/3}Zr_{1/3}Sn_{1/3})O_3$, have the same characteristic of an ideal cubic phase structure, a broad light absorption and almost similar band structure with similar photocatalytic hydrogen production. A main benefit of such HEPs is that they do not need a platinum cocatalyst to produce hydrogen. These results show the potential of HEP materials for cocatalyst-free photocatalytic applications.

**CRediT Authorship Contribution Statement**

All authors: Conceptualization, Methodology, Investigation, Validation, Writing – review & editing.

**Declaration of Competing Interest**

The authors declare no competing financial interests or personal relationships that can affect the research presented in this manuscript.

**Data availability**

All the data presented in this article are made available through the request to the corresponding author

**Acknowledgments**

H.T.N.H. expresses sincere appreciation to the Yoshida Scholarship Foundation (YSF) for providing a Ph.D. scholarship. This research is supported in part by funding from Mitsui Chemicals, Inc., Japan, a Grant-in-Aid from the ASPIRE initiative of the Japan Science and Technology Agency (JST) (JPMJAP2332).

**Reference**

[1] Y. Zheng, M. Ma, H. Shao, Recent advances in efficient and scalable solar hydrogen production through water splitting, Carbon Neutrality 2 (2023) 23. https://doi.org/10.1007/s43979-023-00064-6.

[2] S. Nishioka, F.E. Osterloh, X. Wang, T.E. Mallouk, K. Maeda, Photocatalytic water splitting, Nat. Rev. Methods Primer 3 (2023) 42. https://doi.org/10.1038/s43586-023-00226-x.

[3] J. Herrmann, Photocatalysis, in: Kirk-Othmer (Ed.), Kirk-Othmer Encycl. Chem. Technol., 1st ed., Wiley, 2017: pp. 1–44. https://doi.org/10.1002/0471238961.1608152019051816.a01.pub3.

[4] J. Yang, D. Wang, H. Han, C. Li, Roles of Cocatalysts in Photocatalysis and Photoelectrocatalysis, Acc. Chem. Res. 46 (2013) 1900–1909. https://doi.org/10.1021/ar300227e.

[5] M. Marchelek, M. Diak, M. Kozak, A. Zaleska-Medynska, E. Grabowska, Some Unitary, Binary, and Ternary Non-$TiO_2$ Photocatalysts, in: W. Cao (Ed.), Semicond. Photocatal. - Mater. Mech. Appl., InTech, 2016. https://doi.org/10.5772/62583.

[6] S. Takasugi, K. Tomita, M. Iwaoka, H. Kato, M. Kakihana, The hydrothermal and solvothermal synthesis of $LiTaO_3$ photocatalyst: Suppressing the deterioration of the water splitting activity without using a cocatalyst, Int. J. Hydrog. Energy 40 (2015) 5638–5643. https://doi.org/10.1016/j.ijhydene.2015.02.121.

[7] M. Liu, G. Zhang, X. Liang, Z. Pan, D. Zheng, S. Wang, Z. Yu, Y. Hou, X. Wang, $Rh/Cr_2O_3$ and $CoO_x$ Cocatalysts for Efficient Photocatalytic Water Splitting by Poly (Triazine Imide) Crystals, Angew. Chem. Int. Ed. 62 (2023) e202304694. https://doi.org/10.1002/anie.202304694.

[8] B. Li, W. Wang, J. Zhao, Z. Wang, B. Su, Y. Hou, Z. Ding, W.-J. Ong, S. Wang, All-solid-state direct Z-scheme $NiTiO_3/Cd_{0.5}Zn_{0.5}S$ heterostructures for photocatalytic hydrogen evolution with visible light, J. Mater. Chem. A 9 (2021) 10270–10276. https://doi.org/10.1039/D1TA01220G.

[9] J. Su, J. Zhang, S. Chai, Y. Wang, S. Wang, Y. Fang, Optimizing Poly(heptazine imide) Photoanodes Using Binary Molten Salt Synthesis for Water Oxidation Reaction, Acta Phys.-Chim. Sin. 40 (2024) 2408012. https://doi.org/10.3866/PKU.WHXB202408012.

[10] Q. Wang, D. Zheng, Z. Pan, W. Xing, S. Wang, Y. Hou, M. Anpo, G. Zhang, Spatially Separated Redox Cocatalysts on Poly Triazine Imides Boosting Photocatalytic Overall Water Splitting, Adv. Funct. Mater. 35 (2025) 2501889. https://doi.org/10.1002/adfm.202501889.

[11] J. Hidalgo-Jiménez, T. Akbay, X. Sauvage, T. Ishihara, K. Edalati, Photocatalytic carbon dioxide methanation by high-entropy oxides: Significance of work function, Appl. Catal. B Environ. Energy 371 (2025) 125259. https://doi.org/10.1016/j.apcatb.2025.125259.

[12] X. Tang, Z. Ding, Z. Wang, N. Arif, Y. Chen, L. Li, Y. Zeng, Recent Advances in Photocatalytic Nitrogen Fixation Based on Two-Dimensional Materials, ChemCatChem 16 (2024) e202401355. https://doi.org/10.1002/cctc.202401355.

[13] T.T. Nguyen, J. Hidalgo-Jiménez, X. Sauvage, K. Saito, Q. Guo, K. Edalati, Phase and sulfur vacancy engineering in cadmium sulfide for boosting hydrogen production from catalytic plastic waste photoconversion, Chem. Eng. J. 504 (2025) 158730. https://doi.org/10.1016/j.cej.2024.158730.

[14] H.T.N. Hai, T.T. Nguyen, M. Nishibori, T. Ishihara, K. Edalati, Photoreforming of plastic waste into valuable products and hydrogen using a high-entropy oxynitride with distorted atomic-scale structure, Appl. Catal. B Environ. Energy 365 (2025) 124968. https://doi.org/10.1016/j.apcatb.2024.124968.

[15] D. Gunawan, J. Zhang, Q. Li, C.Y. Toe, J. Scott, M. Antonietti, J. Guo, R. Amal, Materials Advances in Photocatalytic Solar Hydrogen Production: Integrating Systems and Economics for a Sustainable Future, Adv. Mater. (2024) 2404618. https://doi.org/10.1002/adma.202404618.

[16] S.K. Lakhera, A. Rajan, R. T.P., N. Bernaurdshaw, A review on particulate photocatalytic hydrogen production system: Progress made in achieving high energy conversion efficiency and key challenges ahead, Renew. Sustain. Energy Rev. 152 (2021) 111694. https://doi.org/10.1016/j.rser.2021.111694.

[17] X. Du, H. Ji, Y. Xu, S. Du, Z. Feng, B. Dong, R. Wang, F. Zhang, Covalent organic framework without cocatalyst loading for efficient photocatalytic sacrificial hydrogen production from water, Nat. Commun. 16 (2025) 3024. https://doi.org/10.1038/s41467-025-58337-w.

[18] L. Tian, X. Guan, S. Zong, A. Dai, J. Qu, Cocatalysts for Photocatalytic Overall Water Splitting: A Mini Review, Catalysts 13 (2023) 355. https://doi.org/10.3390/catal13020355.

[19] K. Maeda, K. Domen, Photocatalytic Water Splitting: Recent Progress and Future Challenges, J. Phys. Chem. Lett. 1 (2010) 2655–2661. https://doi.org/10.1021/jz1007966.

[20] H.-R. Wenk, A.G. Bulakh, Minerals: their constitution and origin, Cambridge University Press, Cambridge ; New York, 2004.

[21] M. Irshad, Q.T. Ain, M. Zaman, M.Z. Aslam, N. Kousar, M. Asim, M. Rafique, K. Siraj, A.N. Tabish, M. Usman, M.U. Hassan Farooq, M.A. Assiri, M. Imran, Photocatalysis and perovskite oxide-based materials: a remedy for a clean and sustainable future, RSC Adv. 12 (2022) 7009–7039. https://doi.org/10.1039/D1RA08185C.

[22] M.R. Filip, F. Giustino, The geometric blueprint of perovskites, Proc. Natl. Acad. Sci. 115 (2018) 5397–5402. https://doi.org/10.1073/pnas.1719179115.

[23] M.A. Peña, J.L.G. Fierro, Chemical Structures and Performance of Perovskite Oxides, Chem. Rev. 101 (2001) 1981–2018. https://doi.org/10.1021/cr980129f.

[24] J. Ma, T. Liu, W. Ye, Q. He, K. Chen, High-entropy perovskite oxides for energy materials: A review, J. Energy Storage 90 (2024) 111890. https://doi.org/10.1016/j.est.2024.111890.

[25] Y. Wang, M.J. Robson, A. Manzotti, F. Ciucci, High-entropy perovskites materials for next-generation energy applications, Joule 7 (2023) 848–854. https://doi.org/10.1016/j.joule.2023.03.020.

[26] J. Dąbrowa, A. Olszewska, A. Falkenstein, C. Schwab, M. Szymczak, M. Zajusz, M. Moździerz, A. Mikuła, K. Zielińska, K. Berent, T. Czeppe, M. Martin, K. Świerczek, An innovative approach to design SOFC air electrode materials: high entropy $La_{1-x}Sr_x(Co,Cr,Fe,Mn,Ni)O_{3-\delta}$ ($x = 0, 0.1, 0.2, 0.3$) perovskites synthesized by the sol–gel method, J. Mater. Chem. A 8 (2020) 24455–24468. https://doi.org/10.1039/D0TA06356H.

[27] S. Jiang, T. Hu, J. Gild, N. Zhou, J. Nie, M. Qin, T. Harrington, K. Vecchio, J. Luo, A new class of high-entropy perovskite oxides, Scr. Mater. 142 (2018) 116–120. https://doi.org/10.1016/j.scriptamat.2017.08.040.

[28] W. Sun, F. Zhang, X. Zhang, T. Shi, J. Li, Y. Bai, C. Wang, Z. Wang, Enhanced electrical properties of $(Bi_{0.2}Na_{0.2}Ba_{0.2}Ca_{0.2}Sr_{0.2})TiO_3$ high-entropy ceramics prepared by hydrothermal method, Ceram. Int. 48 (2022) 19492–19500. https://doi.org/10.1016/j.ceramint.2022.04.139.

[29] T.X. Nguyen, Y. Liao, C. Lin, Y. Su, J. Ting, Advanced High Entropy Perovskite Oxide Electrocatalyst for Oxygen Evolution Reaction, Adv. Funct. Mater. 31 (2021) 2101632. https://doi.org/10.1002/adfm.202101632.

[30] M. Biesuz, S. Fu, J. Dong, A. Jiang, D. Ke, Q. Xu, D. Zhu, M. Bortolotti, M.J. Reece, C. Hu, S. Grasso, High entropy $Sr((Zr_{0.94}Y_{0.06})_{0.2}Sn_{0.2}Ti_{0.2}Hf_{0.2}Mn_{0.2})O_{3-x}$ perovskite synthesis by reactive spark plasma sintering, J. Asian Ceram. Soc. 7 (2019) 127–132. https://doi.org/10.1080/21870764.2019.1595931.

[31] A. Farhan, F. Stramaglia, M. Cocconcelli, N. Kuznetsov, L. Yao, A. Kleibert, C. Piamonteze, S. Van Dijken, Weak ferromagnetism in $Tb(Fe_{0.2}Mn_{0.2}Co_{0.2}Cr_{0.2}Ni_{0.2})O_3$ high-entropy oxide perovskite thin films, Phys. Rev. B 106 (2022) L060404. https://doi.org/10.1103/PhysRevB.106.L060404.

[32] K. Edalati, Z. Horita, A review on high-pressure torsion (HPT) from 1935 to 1988, Mater. Sci. Eng. A 652 (2016) 325–352. https://doi.org/10.1016/j.msea.2015.11.074.

[33] L. Tang, Z. Li, K. Chen, C. Li, X. Zhang, L. An, High-entropy oxides based on valence combinations: design and practice, J. Am. Ceram. Soc. 104 (2021) 1953–1958. https://doi.org/10.1111/jace.17659.

[34] B. Charles, J. Dillon, O.J. Weber, M.S. Islam, M.T. Weller, Understanding the stability of mixed A-cation lead iodide perovskites, J. Mater. Chem. A 5 (2017) 22495–22499. https://doi.org/10.1039/c7ta08617b.

[35] N. Ashurov, B.L. Oksengendler, S. Maksimov, S. Rashiodva, A.R. Ishteev, D.S. Saranin, I.N. Burmistrov, D.V. Kuznetsov, A.A. Zakhisov, Current state and perspectives for organo-halide perovskite solar cells. Part 1. Crystal structures and thin film formation, morphology, processing, degradation, stability improvement by carbon nanotubes. A review, Mod. Electron. Mater. 3 (2017) 1–25. https://doi.org/10.1016/j.moem.2017.05.001.

[36] K. Djebari, A. Dahani, M. Djermouni, K. Dine, A. Cherifi, O. Arbouche, A. Zaoui, S. Kacimi, Spontaneous polarization study in $A^{3+}B^{4+}(O_2N)^{7-}$ and $A^{2+}B^{5+}(O_2N)^{7-}$ perovskite-type oxynitrides: a first principles investigation, Appl. Phys. A 128 (2022). https://doi.org/10.1007/s00339-022-05535-8.

[37] W. Li, E. Ionescu, R. Riedel, A. Gurlo, Can we predict the formability of perovskite oxynitrides from tolerance and octahedral factors?, J. Mater. Chem. A 1 (2013) 12239. https://doi.org/10.1039/c3ta10216e.

[38] D. Saikia, A. Betal, J. Bera, S. Sahu, Progress and challenges of halide perovskite-based solar cell- a brief review, Mater. Sci. Semicond. Process. 150 (2022) 106953. https://doi.org/10.1016/j.mssp.2022.106953.

[39] Z. Gao, G. Mao, S. Chen, Y. Bai, P. Gao, C. Wu, I.D. Gates, W. Yang, X. Ding, J. Yao, High throughput screening of promising lead-free inorganic halide double perovskites *via* first-principles calculations, Phys. Chem. Chem. Phys. 24 (2022) 3460–3469. https://doi.org/10.1039/d1cp04976c.

[40] M. Mittal, R. Garg, A. Jana, Recent progress in the stabilization of low band-gap black-phase iodide perovskite solar cells, Dalton Trans. 52 (2023) 11750–11767. https://doi.org/10.1039/d3dt01581e.

[41] J. Hidalgo-Jiménez, T. Akbay, X. Sauvage, L. Van Eijck, M. Watanabe, J. Huot, T. Ishihara, K. Edalati, Hybrid $d^0$ and $d^{10}$ electronic configurations promote photocatalytic activity of high-entropy oxides for $CO_2$ conversion and water splitting, J. Mater. Chem. A 12 (2024) 31589–31602. https://doi.org/10.1039/d4ta04689g.

[42] J. Hidalgo-Jiménez, T. Akbay, X. Sauvage, T. Ishihara, K. Edalati, Mixed atomic-scale electronic configuration as a strategy to avoid cocatalyst utilization in photocatalysis by high-entropy oxides, Acta Mater. 283 (2025) 120559. https://doi.org/10.1016/j.actamat.2024.120559.

[43] C.J. Bartel, C. Sutton, B.R. Goldsmith, R. Ouyang, C.B. Musgrave, L.M. Ghiringhelli, M. Scheffler, New tolerance factor to predict the stability of perovskite oxides and halides, Sci. Adv. 5 (2019) eaav0693. https://doi.org/10.1126/sciadv.aav0693.

[44] K.F. Moura, L. Chantelle, D. Rosendo, E. Longo, I.M.G.D. Santos, Effect of $Fe^{3+}$ Doping in the Photocatalytic Properties of $BaSnO_3$ Perovskite, Mater. Res. 20 (2017) 317–324. https://doi.org/10.1590/1980-5373-mr-2016-1062.

[45] V.K. Veerapandiyan, S. Khosravi H, G. Canu, A. Feteira, V. Buscaglia, K. Reichmann, M. Deluca, B-site vacancy induced Raman scattering in $BaTiO_3$-based ferroelectric ceramics, J. Eur. Ceram. Soc. 40 (2020) 4684–4688. https://doi.org/10.1016/j.jeurceramsoc.2020.05.051.

[46] V.S. Puli, P. Li, S. Adireddy, D.B. Chrisey, Crystal structure, dielectric, ferroelectric and energy storage properties of La-doped $BaTiO_3$ semiconducting ceramics, J. Adv. Dielectr. 05 (2015) 1550027. https://doi.org/10.1142/S2010135X15500277.

[47] L. Veselinović, M. Mitrić, L. Mančić, M. Vukomanović, B. Hadžić, S. Marković, D. Uskoković, The effect of Sn for Ti substitution on the average and local crystal structure of $BaTi_{1-x}Sn_xO_3$ ($0 \leq x \leq 0.20$), J. Appl. Crystallogr. 47 (2014) 999–1007. https://doi.org/10.1107/S1600576714007584.

[48] S. Sumithra, N.V. Jaya, Tunable Optical Behaviour and Room Temperature Ferromagnetism of Cobalt-Doped $BaSnO_3$ Nanostructures, J. Supercond. Nov. Magn. 31 (2018) 2777–2787. https://doi.org/10.1007/s10948-017-4504-8.

[49] L. Yuan, C. Xu, S. Zhang, M. Yu, X. Wang, Y. Chen, L. Dai, Role of oxygen vacancy in spinel $(FeCoNiCrMn)_3O_4$ high entropy oxides prepared via two different methods for the selective C H bond oxidation of p-chlorotoluene, J. Colloid Interface Sci. 640 (2023) 359–371. https://doi.org/10.1016/j.jcis.2023.02.128.

[50] C. Wang, W. Liu, M. Liao, J. Weng, J. Shen, Y. Chen, Y. Du, Novel nano spinel-type high-entropy oxide (HEO) catalyst for hydrogen production using ethanol steam reforming, Nanoscale 15 (2023) 8619–8632. https://doi.org/10.1039/D2NR07195A.

[51] K. Wenderich, G. Mul, Methods, Mechanism, and Applications of Photodeposition in Photocatalysis: A Review, Chem. Rev. 116 (2016) 14587–14619. https://doi.org/10.1021/acs.chemrev.6b00327.

[52] S. Assavachin, F.E. Osterloh, Ferroelectric Polarization in $BaTiO_3$ Nanocrystals Controls Photoelectrochemical Water Oxidation and Photocatalytic Hydrogen Evolution, J. Am. Chem. Soc. 145 (2023) 18825–18833. https://doi.org/10.1021/jacs.3c03762.

[53] A.P. Souri, E. Skliri, I. Vamvasakis, G.S. Armatas, V. Binas, Highly active pt nanoparticles supported on $SrTiO_3$ for photocatalytic hydrogen production, Appl. Phys. A 130 (2024) 785. https://doi.org/10.1007/s00339-024-07967-w.

[54] S.N. Tijare, M.V. Joshi, P.S. Padole, P.A. Mangrulkar, S.S. Rayalu, N.K. Labhsetwar, Photocatalytic hydrogen generation through water splitting on nano-crystalline $LaFeO_3$ perovskite, Int. J. Hydrog. Energy 37 (2012) 10451–10456. https://doi.org/10.1016/j.ijhydene.2012.01.120.

[55] R. Jacobs, J. Hwang, Y. Shao-Horn, D. Morgan, Assessing Correlations of Perovskite Catalytic Performance with Electronic Structure Descriptors, Chem. Mater. 31 (2019) 785–797. https://doi.org/10.1021/acs.chemmater.8b03840.

[56] M. Humayun, Z. Li, M. Israr, A. Khan, W. Luo, C. Wang, Z. Shao, Perovskite Type $ABO_3$ Oxides in Photocatalysis, Electrocatalysis, and Solid Oxide Fuel Cells: State of the Art and Future Prospects, Chem. Rev. 125 (2025) 3165–3241. https://doi.org/10.1021/acs.chemrev.4c00553.

[57] J.A. Perez Franco, A. García Murillo, F.D.J. Carrillo Romo, I.C. Romero Ibarra, A. Cervantes Tobón, Identifying Crystal Structure of Halides of Strontium and Barium Perovskite Compounds with EXPO2014 Software, Materials 18 (2024) 58. https://doi.org/10.3390/ma18010058.

[58] W.D. Mesquita, S.R. De Jesus, M.C. Oliveira, R.A.P. Ribeiro, M.R. De Cássia Santos, M.G. Junior, E. Longo, M.F. Do Carmo Gurgel, Barium strontium titanate-based perovskite materials from DFT perspective: assessing the structural, electronic, vibrational, dielectric and energetic properties, Theor. Chem. Acc. 140 (2021) 27. https://doi.org/10.1007/s00214-021-02723-2.

[59] W. Li, K. Jiang, Z. Li, S. Gong, R.L.Z. Hoye, Z. Hu, Y. Song, C. Tian, J. Kim, K.H.L. Zhang, S. Cho, J.L. MacManus-Driscoll, Origin of Improved Photoelectrochemical Water Splitting in Mixed Perovskite Oxides, Adv. Energy Mater. 8 (2018). https://doi.org/10.1002/aenm.201801972.

[60] M. Maqbool, B. Amin, I. Ahmad, Bandgap investigations and the effect of the In and Al concentration on the optical properties of $In_xAl_{1-x}N$, J. Opt. Soc. Am. B 26 (2009) 2181. https://doi.org/10.1364/JOSAB.26.002181.

[61] Y. Zhang, J. Tian, D. Tian, W. Li, Z. Liu, F. Tian, Y. Bu, S. Kuang, Regulation of band edge and specific surface area of $Bi_xIn_yOCl$ microsphere for excellent photocatalytic performance, J. Alloys Compd. 867 (2021) 159052. https://doi.org/10.1016/j.jallcom.2021.159052.

[62] K. Edalati, K. Fujiwara, S. Takechi, Q. Wang, M. Arita, M. Watanabe, X. Sauvage, T. Ishihara, Z. Horita, Improved Photocatalytic Hydrogen Evolution on Tantalate Perovskites $CsTaO_3$ and $LiTaO_3$ by Strain-Induced Vacancies, ACS Appl. Energy Mater. 3 (2020) 1710–1718. https://doi.org/10.1021/acsaem.9b02197.

[63] Y. Hu, L. Mao, X. Guan, K.A. Tucker, H. Xie, X. Wu, J. Shi, Layered perovskite oxides and their derivative nanosheets adopting different modification strategies towards better photocatalytic performance of water splitting, Renew. Sustain. Energy Rev. 119 (2020) 109527. https://doi.org/10.1016/j.rser.2019.109527.

[64] Y. Wei, J. Wang, R. Yu, J. Wan, D. Wang, Constructing SrTiO$_3$ –TiO$_2$ Heterogeneous Hollow Multi-shelled Structures for Enhanced Solar Water Splitting, Angew. Chem. 131 (2019) 1436–1440. https://doi.org/10.1002/ange.201812364.

[65] Y. Yuan, Z. Zhao, J. Zheng, M. Yang, L. Qiu, Z. Li, Z. Zou, Polymerizable complex synthesis of BaZr$_{1-x}$Sn$_x$O$_3$ photocatalysts: Role of Sn$^{4+}$ in the band structure and their photocatalytic water splitting activities, J. Mater. Chem. 20 (2010) 6772. https://doi.org/10.1039/c0jm00455c.

[66] C.-C. Hu, Y.-L. Lee, H. Teng, Efficient water splitting over Na$_{1-x}$K$_x$TaO$_3$ photocatalysts with cubic perovskite structure, J. Mater. Chem. 21 (2011) 3824. https://doi.org/10.1039/c0jm03451g.

[67] Z. Zhao, J. Wu, Y.-Z. Zheng, N. Li, X. Li, Z. Ye, S. Lu, X. Tao, C. Chen, Stable hybrid perovskite MAPb(I$_{1-x}$Br$_x$)3 for photocatalytic hydrogen evolution, Appl. Catal. B Environ. 253 (2019) 41–48. https://doi.org/10.1016/j.apcatb.2019.04.050.

[68] H.T.N. Hai, J. Hidalgo-Jiménez, K. Edalati, Boosting hydrogen and methane formation on a high-entropy photocatalyst by integrating atomic d$^0$/d$^{10}$ electronic junctions and microscopic P/N heterojunctions, Int. J. Hydrog. Energy 162 (2025) 150762. https://doi.org/10.1016/j.ijhydene.2025.150762.

[69] T.T. Nguyen, J. Hidalgo-Jiménez, K. Edalati, High-entropy oxide photocatalysts for green ammonia synthesis from nitrogen fixation in water, Scr. Mater. 269 (2025) 116931. https://doi.org/10.1016/j.scriptamat.2025.116931.

[70] T.T. Nguyen, K. Edalati, High-entropy oxide with tailored heterogeneous electronic structure as a low-bandgap catalyst for antibiotic photodegradation under visible light, Appl. Catal. B Environ. Energy 382 (2026) 126011. https://doi.org/10.1016/j.apcatb.2025.126011.

[71] P. Edalati, Y. Itagoe, H. Ishihara, T. Ishihara, H. Emami, M. Arita, M. Fuji, K. Edalati, Visible-light photocatalytic oxygen production on a high-entropy oxide with multiple-heterojunction introduction, J. Photochem. Photobio. A 433 (2022) 114167. https://doi.org/10.1016/j.jphotochem.2022.114167.